\documentclass{DISproc}

\begin{document}
%------------------------------------
\title{Summary of the Electroweak and Searches\\ Working Group}

%for single authors the superscripts are optional
\author{{\slshape David M. South$^1$, Andreas Weiler$^1$, Hwidong Yoo$^2$}\\[1ex]
$^1$DESY, Notkestra{\ss}e 85, 22607 Hamburg, Germany\\
$^2$Purdue University, West Lafayette, IN 47907, USA }

% please enter the contribution ID for the DOI
\contribID{359}

\doi  % if there is an online version we will register DOIs

\maketitle

\begin{abstract}
The working group on electroweak measurements and searches for new physics at the Deep Inelastic Scattering
2012 workshop covered a wide range of results from the various experiments at the LHC (ATLAS, CMS, LHCb),
the Tevatron (CDF, D{\O}) and HERA (H1, ZEUS), as well as results from the BaBar, NA48/62 and
OPERA collaborations.
In addition, invited theoretical overviews were presented and discussed in each of the sessions.
A summary of a selection of the results shown at the conference is given.
\end{abstract}

\section{Introduction}
\label{sec:intro}

The electroweak and searches working group at the Deep Inelastic Scattering 2012 conference
included some of the most exciting results yet in particle physics.
The phenomenal performance of the LHC in 2011 allowed substantial $pp$ data sets to be analysed
at a centre-of-mass energy $\sqrt{s}$ beyond what has previously been explored.
In addition, many of the results shown at the conference by the ATLAS and CMS collaborations
included the full $5$~fb$^{-1}$ of 2011 data taken at $\sqrt{s}=7$~TeV,  an impressive effort.
Complementing the data from the LHC, are the $p\bar{p}$ collisions at the Tevatron, where
results are now emerging using the full $10$~fb$^{-1}$ Run-II data sets from the CDF and
D{\O} collaborations.
By utilising as much data as possible, the reach of searches has been significantly extended
and the precision of Standard Model (SM) electroweak measurements has been greatly
enhanced\footnote{Note that these proceedings summarise the results available at the time
of the conference and more recent results, in particular in the case of Higgs analyses,
may be available.}.
The final searches in $e^{\pm}p$ data from the HERA collaborations, as well as results from
the BaBar, NA48/62 and OPERA collaborations, completed the picture.
On the final day of the conference, collisions began again at the LHC, at an increased centre of mass
energy of $8$~TeV.
The forthcoming analysis of this data will provide further insight into searches
for physics beyond the SM and precision electroweak physics.

\section{Standard Model Higgs searches}
\label{sec:smhiggs}

The search for the SM Higgs boson represents one of the most important endeavours of modern
particle physics.
The dominant Higgs production mechanism at hadron colliders is via gluon fusion, where other
processes - associated vector boson production (VH), vector boson fusion (VBF) and top fusion -
are expected to occur at a significantly lower rate.
A review of the current theoretical predictions was presented at the conference~\cite{herzog},
where in particular the variation in the latest PDF and scale uncertainties was discussed.

\begin{figure}
  \centering
  \includegraphics[width=0.51\textwidth]{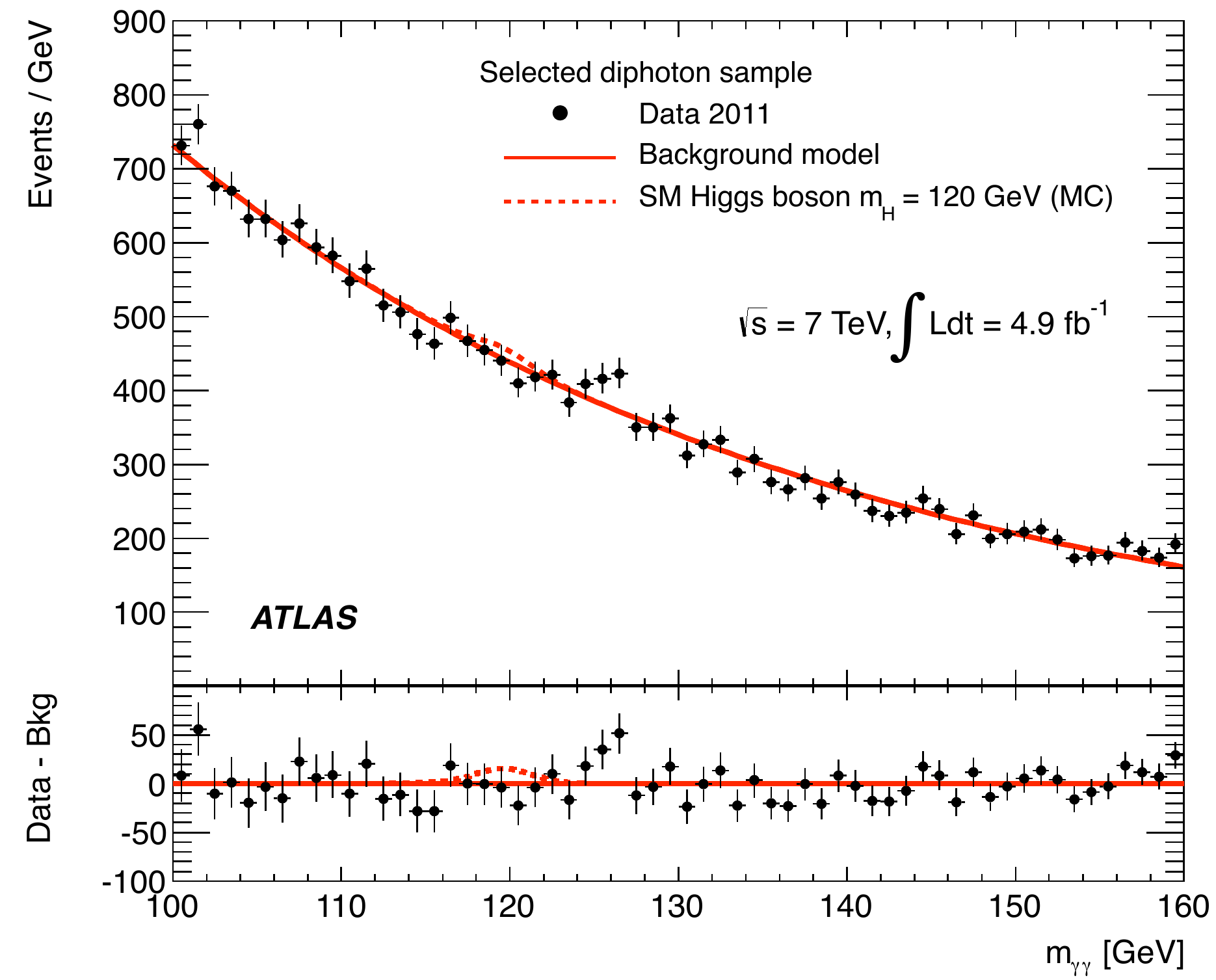}
  \includegraphics[width=0.46\textwidth]{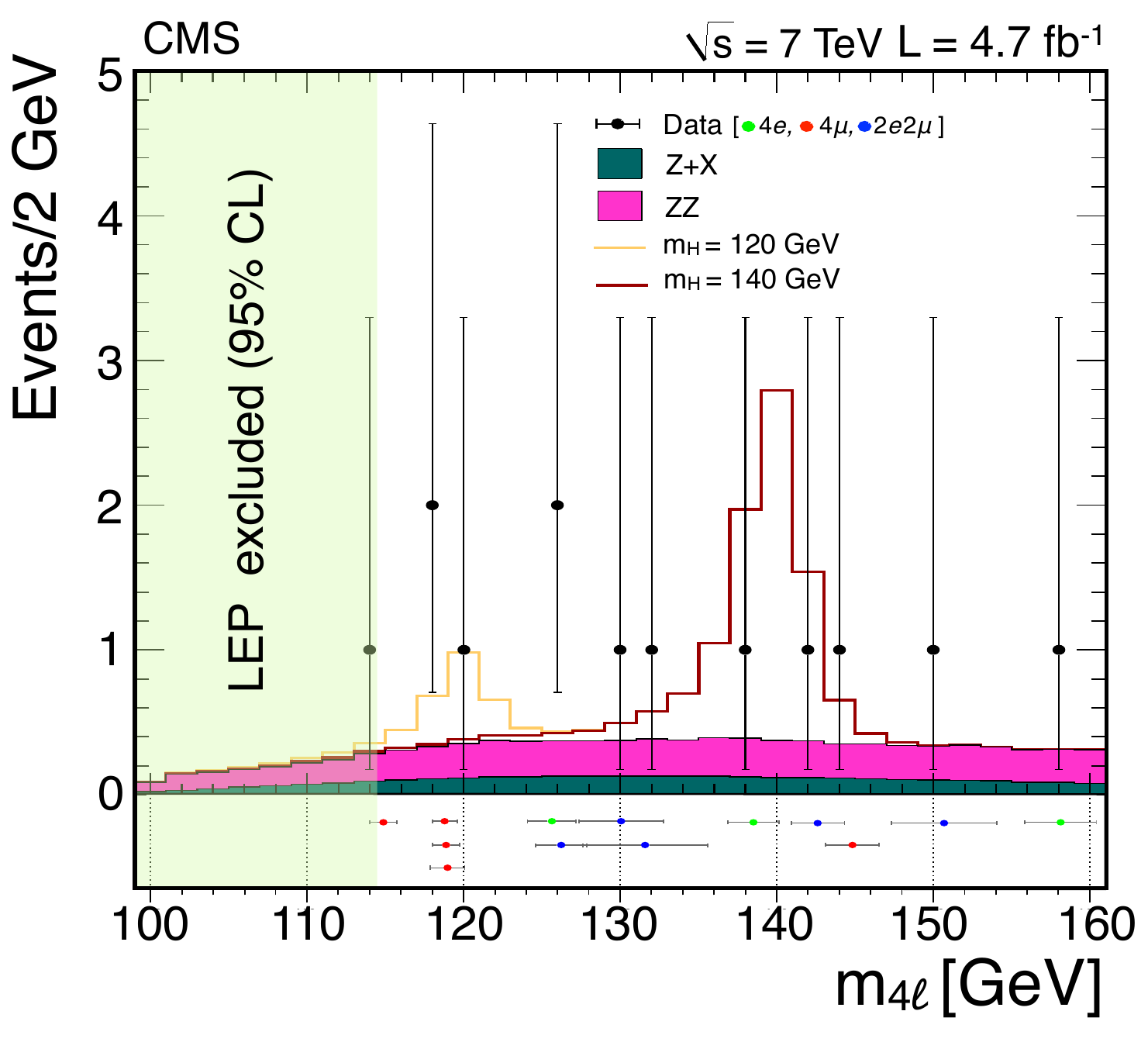}
  \caption{Left: The invariant mass distribution of the diphoton sample in the ATLAS search
    for $H \rightarrow \gamma\gamma$ events, where the Higgs boson expectation for a mass
    hypothesis of  $120$~GeV corresponding to the SM cross section is also shown. Right: The
    $4$-lepton invariant mass distribution in the low mass region of the CMS search for
    $H \rightarrow ZZ \rightarrow 4\ell$ events. The central values and event-by-event mass
    measurement uncertainties are indicated, as well as the expected Higgs boson signal for
    two different mass hypotheses.}
  \label{fig:higgsMassSpectra}
\end{figure}

The SM Higgs boson decays via a variety of channels, where at low masses the
$H \rightarrow b\bar{b}$ and $H \rightarrow \tau\tau$ decays are dominant and for Higgs masses
$M_{H}>135$~GeV the decays to vector boson pairs,  $H \rightarrow ZZ$ and $H \rightarrow WW$
are favoured.
The full spectrum of Higgs decays is investigated by dedicated ATLAS~\cite{mal} and
CMS~\cite{govoni} analyses, which often include different sub-channels, for example of the
different subsequent decays of the vector bosons produced in the Higgs decay.
At the LHC, the two most important channels are $H \rightarrow \gamma\gamma$, which has
a relatively small branching ratio but a clean final state signature, and
$H \rightarrow ZZ \rightarrow 4\ell$, where the final state contains two high-mass pairs of
isolated leptons and has very little SM background.
Both of these channels have good mass resolution and the invariant mass distributions in the ATLAS
diphoton analysis~\cite{ATLAS:2012ad} and the CMS $H \rightarrow ZZ \rightarrow 4\ell$
analysis~\cite{Chatrchyan:2012dg} are shown in Figure~\ref{fig:higgsMassSpectra}, where a small
excess of data events can be seen in both channels between $120$~GeV and $130$~GeV.

\begin{figure}
 \centering
 \includegraphics[width=0.45\textwidth]{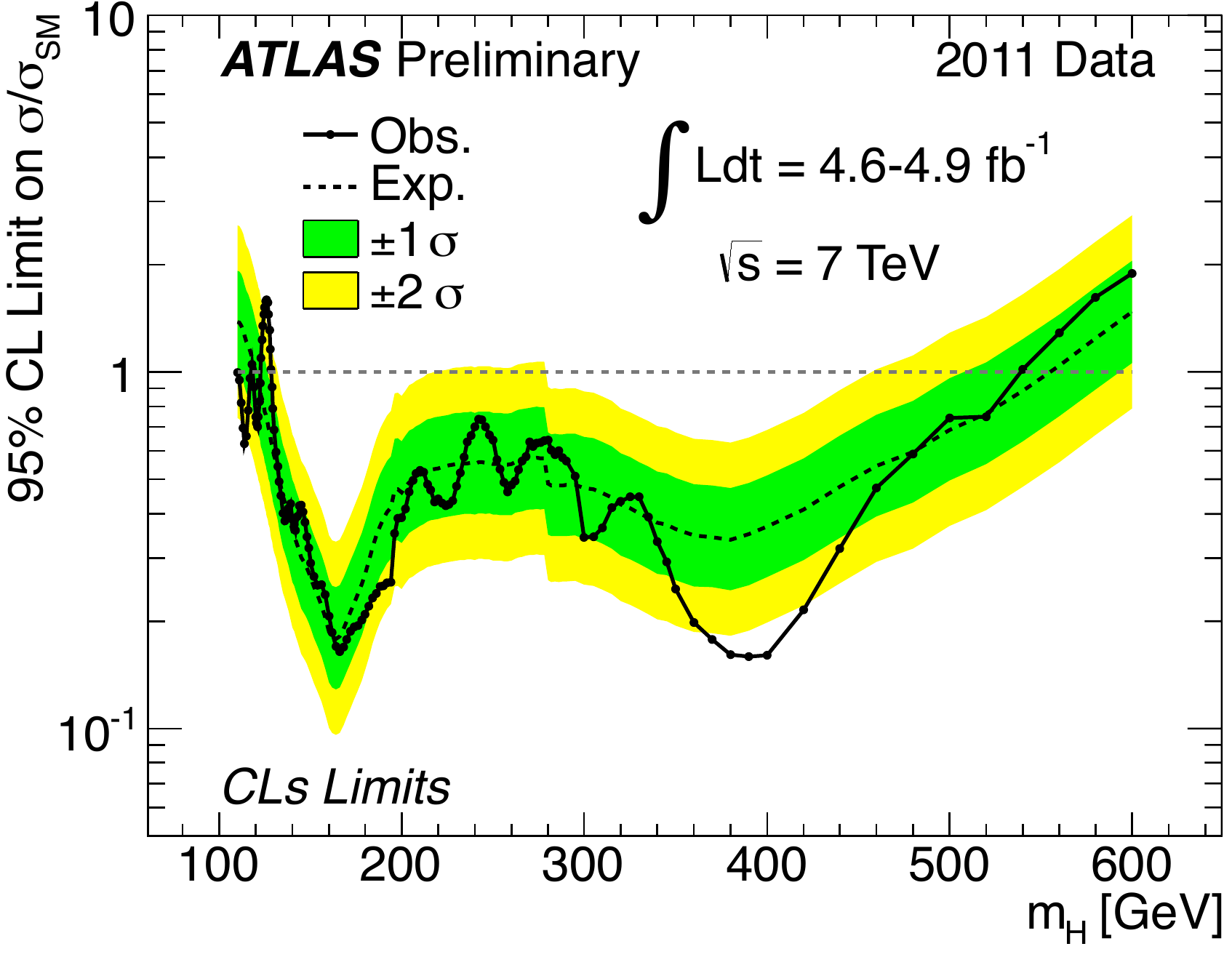} 
 \includegraphics[width=0.52\textwidth]{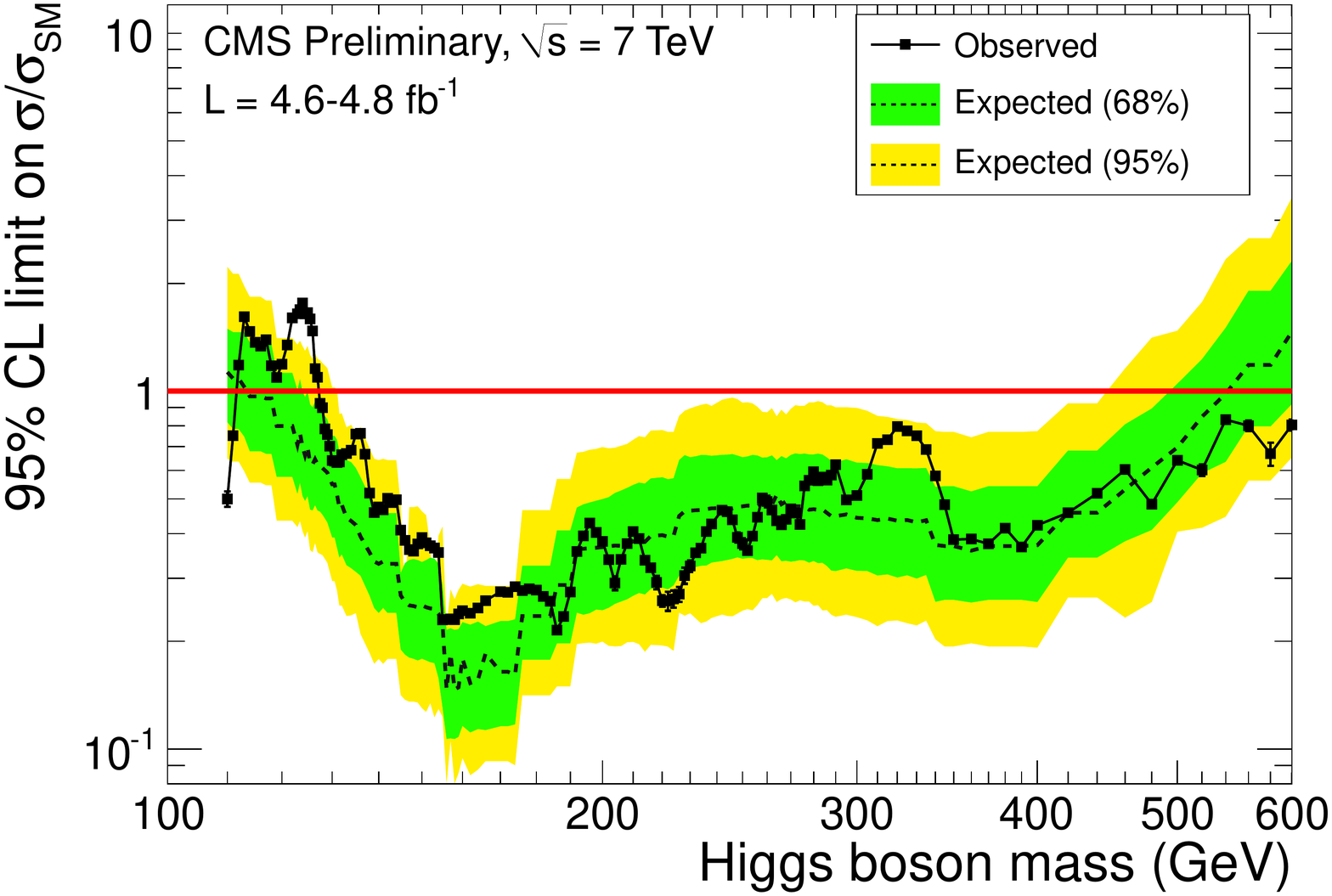}
 \caption{Combined exclusion limits from the ATLAS (left) and CMS (right) searches for the SM
   Higgs boson. The observed $95$\% C.L. combined upper limits on the production cross section
   divided by the SM expectation are shown by the solid line as a function of the Higgs
   mass. The dashed line shows the median expected limit in the absence of a signal and the green
   and yellow bands indicate the corresponding $68$\% and $95$\% intervals.}
 \label{fig:higgsExclusionLimits}
\end{figure}

Combining all investigated decay channels, and using up to their complete 2011 $\sqrt{s}=7$~TeV
data sets, exclusion limits are derived as a function of the Higgs mass by ATLAS~\cite{atlasMoriondHiggs}
and CMS~\cite{cmsMoriondHiggs}, as shown in Figure~\ref{fig:higgsExclusionLimits}.
Higgs boson mass ranges of $110$~GeV to $117.5$~GeV, $118.5$~GeV to $122.5$~GeV and
$129$~GeV to $539$~GeV ($127.5$~GeV to $600$~GeV) are excluded by the ATLAS (CMS) analysis at
the $95$\% confidence level (C.L.), whereas the range $120$~GeV to $553$~GeV ($114.5$~GeV to
$543$~GeV) is expected to be excluded by the ATLAS (CMS) analysis in the absence of a signal. 
The local significance of the observed excess at $126$~GeV in the ATLAS combined analysis
is $2.5\sigma$, rising to $3.5\sigma$ if only the high mass resolution
$H \rightarrow \gamma\gamma$ and $H \rightarrow ZZ \rightarrow 4\ell$ channels are considered.
In the CMS combined analysis, an excess is observed at $124$~GeV with a local significance
of $2.8\sigma$.
In both experiments, the significance of the excess is considerably lower when the full mass range
is taken into account, the so called ``look elsewhere effect''.

The focus of Higgs searches at the Tevatron is on the low mass region, $M_{H} < 135$~GeV,
concentrating on associated vector boson ($W$,$Z$) production with subsequent
$H \rightarrow b\bar{b}$ decays.
Final states are categorised according to the number of charged leptons and the presence or absence
of missing transverse energy (MET) in the final state, arising from the decay of the vector boson.
\begin{wrapfigure}{l}{0.5\textwidth}
  \centering
  \includegraphics[width=0.48\textwidth]{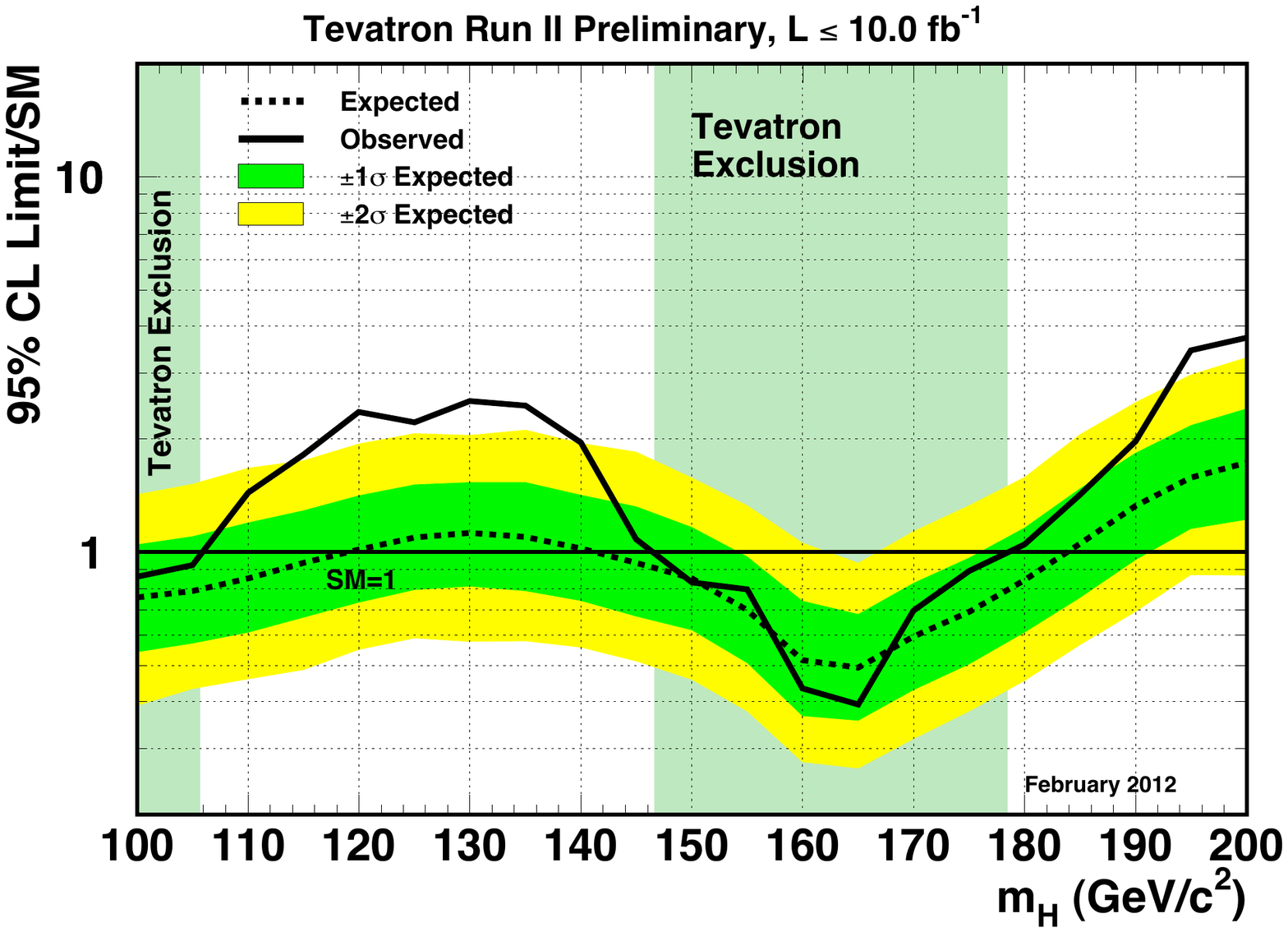}
  \caption{Observed and expected $95$\% C.L. upper limits on the production cross section
    divided by the SM expectation as a function of the Higgs mass for the combined CDF and D{\O}
    analyses. The dotted curves show the median expected limit in the absence of a signal and the
    green and yellow bands indicate the corresponding $68$\% and $95$\% intervals.}
  \label{fig:higgsTevatron}
\end{wrapfigure}
The amount of luminosity analysed by the CDF and D{\O} experiments has increased, where up
to $10$~fb$^{-1}$ from each experiment is now included, and several analysis improvements have
been implemented~\cite{knoepfel}.  
Limits on Higgs production at the Tevatron are shown as a function of mass in
Figure~\ref{fig:higgsTevatron}, where the regions $100$~GeV to $106$~GeV and $147$~GeV to
$179$~GeV are excluded at the $95$\% C.L., compared to the expected exclusion ranges of
$100$~GeV to $119$~GeV and $141$~GeV to $184$~GeV.
In addition, an excess in the data with respect to the background estimation is observed in the
Tevatron analysis across a broad region, which is in the same mass range as the excess observed
by the LHC experiments described above.
The local (global) significance of this excess is $2.7\sigma$ ($2.2\sigma$).
Comparing the observed limits in the presented results from the LHC and the Tevatron, only the
mass ranges $117.5 < M_{H} < 118.5$~GeV and $122.5 < M_{H} < 127.5$~GeV are currently not
excluded at the $95$\% C.L.

\clearpage

\section{Beyond the Standard Model Higgs searches}
\label{sec:bsmhiggs}

Many models of beyond the Standard Model (BSM) physics predict Higgs-like bosons, where the coupling
may be different to that of the SM Higgs, or there may be additional, neutral or charged scalars.
An intriguing possibility are composite Higgs models~\cite{Azatov:2012bz}, where the Higgs is a
bound state of new strong dynamics close to the weak scale~\cite{azatov}.
The Higgs is lighter than the rest of the strong states due to it being a Goldstone boson, as in the
AdS/CFT inspired holographic Higgs or in little Higgs models.
In such a model, new physics is expected to be be revealed via modifications to the SM Higgs couplings.
Many searches for a BSM Higgs are performed at the LHC~\cite{fernandez,lenzi},
the Tevatron~\cite{chapon} and BaBar~\cite{santoro}, and the results of two such searches are
presented in the following.

%%%

\begin{wrapfigure}{r}{0.60\textwidth}
 \centering
 \includegraphics[width=0.58\textwidth]{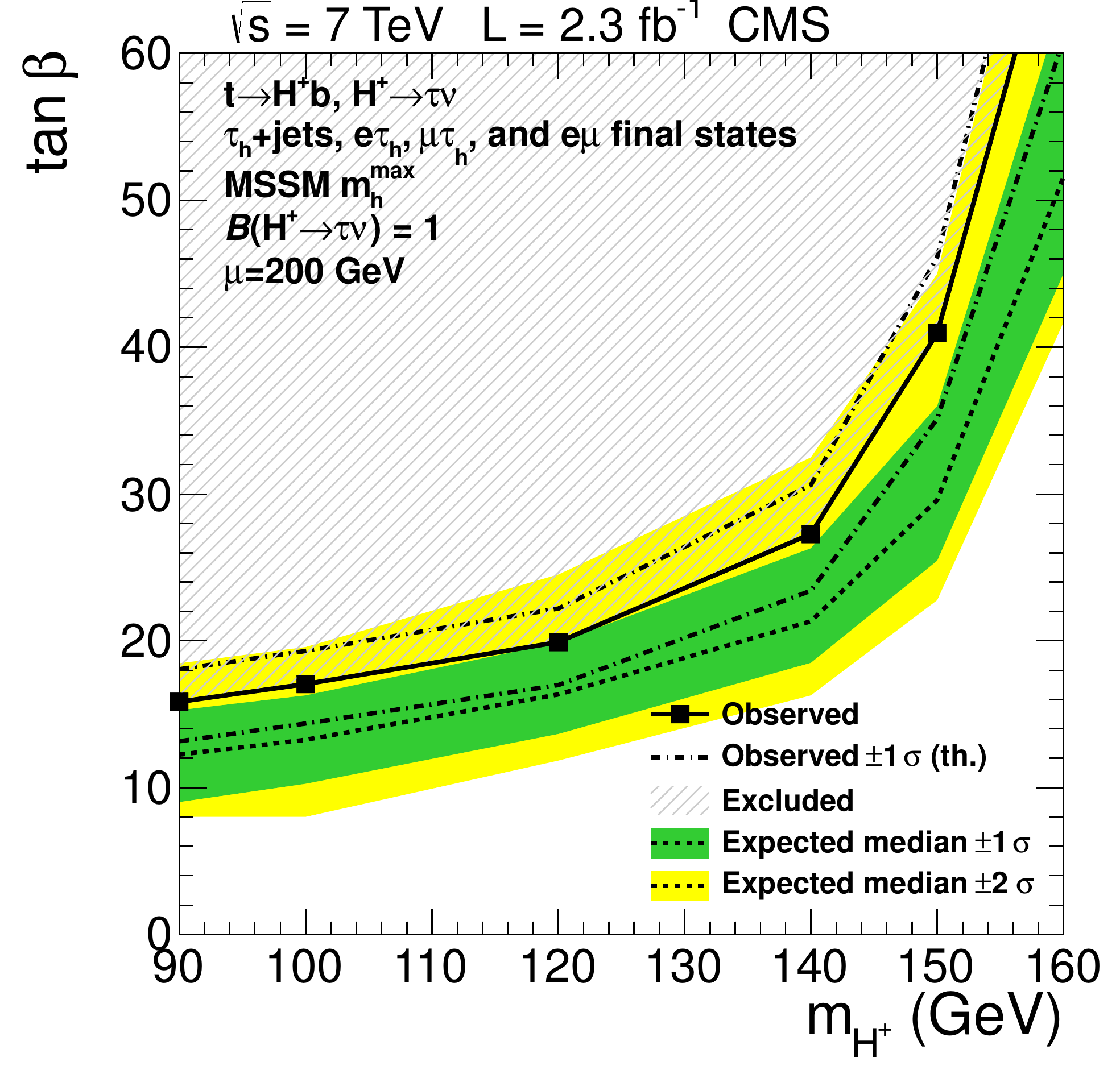}
 \caption{The exclusion region in the $M_{H^{+}}-\tan\beta$  MSSM parameter space obtained from
   the combined CMS analysis for the $M_{H^{+}}^{\rm max}$ MSSM scenario. The green
   and yellow bands indicate the corresponding $68$\% and $95$\% intervals.}
 \label{fig:chargedHiggs}
\end{wrapfigure}
Charged Higgs bosons $H^{\pm}$ are predicted by Higgs doublet models such as the
Minimal Supersymmetric Standard Model (MSSM), as well as Higgs triplet models.
For Higgs masses $M_{H^{\pm}} < M_{t}$ the dominant production mode is via top quark decay
$t\bar{t} \rightarrow b\bar{b}WH^{+}$ and for $\tan\beta >3$ the preferred Higgs decay mode is
$H^{\pm} \rightarrow \tau^{\pm}\nu$.
Searches for charged Higgs bosons are performed by the ATLAS~\cite{atlasChargedHiggs} and
CMS~\cite{cmsChargedHiggs} collaborations, where up to four different final states are analysed
according to combinations of the hadronic and leptonic decays of the $W$ and $H^{+}$.
Good agreement of the data with the SM is observed and in the absence of a signal limits are derived
on the branching ratio $t \rightarrow H^{+}b$.
Limits in the $M_{H^{+}}-\tan\beta$ plane from the CMS analysis are shown in
Figure~\ref{fig:chargedHiggs} for the combination of all examined final states.

%%%

\begin{figure}[t]
 \centering
 \includegraphics[width=0.42\textwidth]{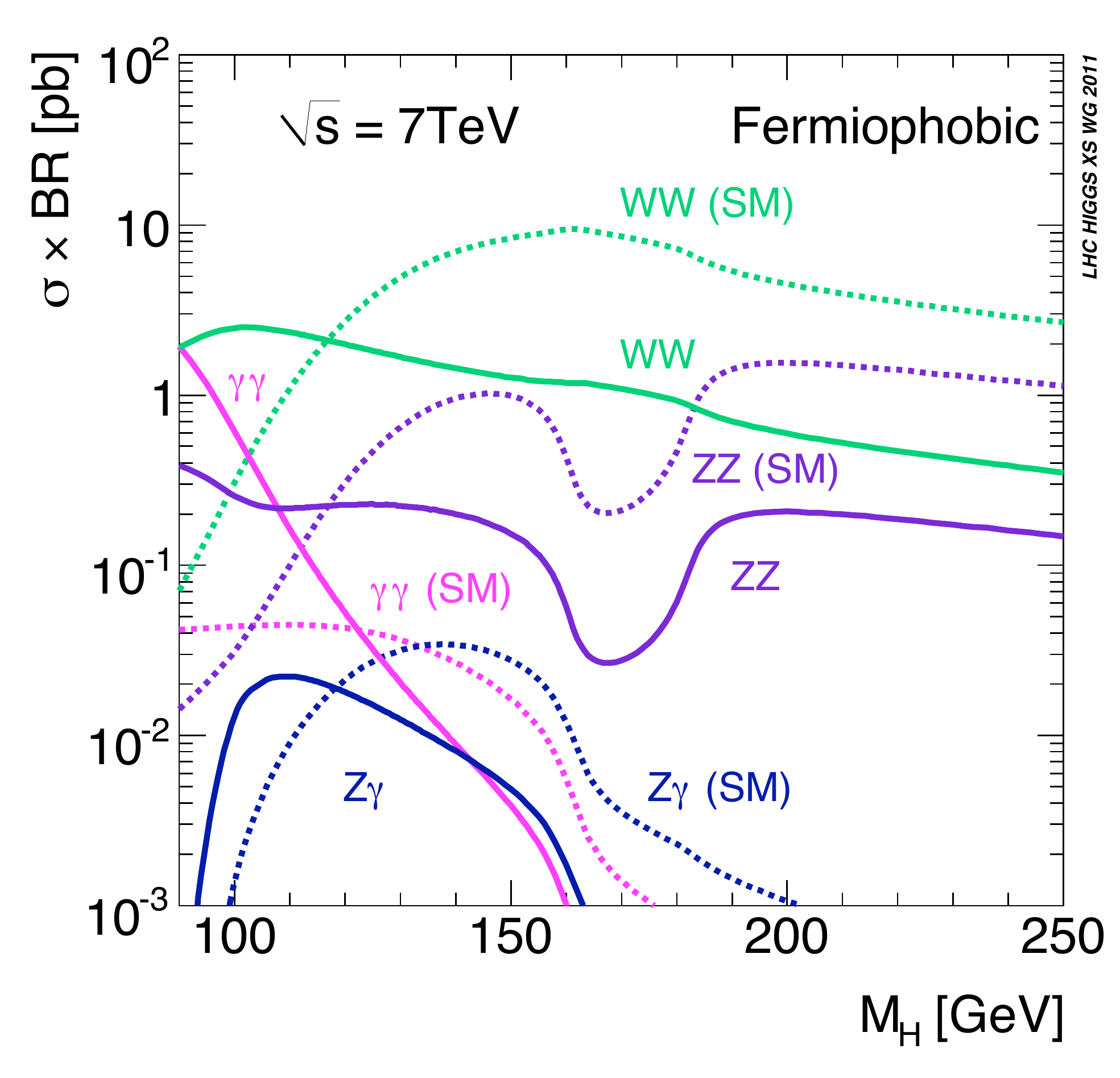}
 \includegraphics[width=0.54\textwidth]{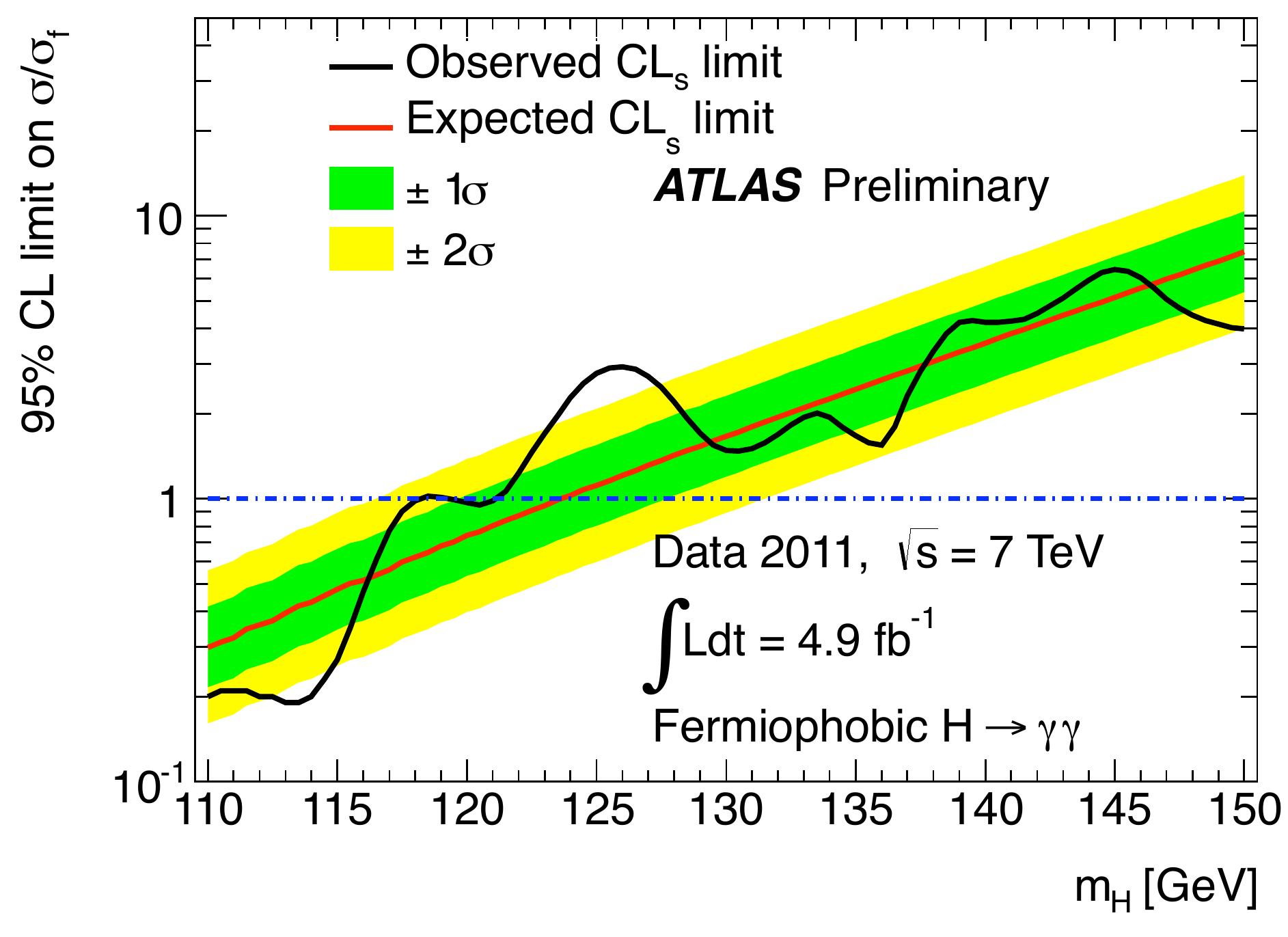}
 \caption{Left: SM (dashed lines) and fermiophobic (solid lines) Higgs production cross section $\times$
   branching ratios for $\sqrt{s}=7$~TeV as a function of the Higgs mass. Right: Observed (black line) and
   expected (red line) $95$\% C.L. limits from the ATLAS experiment for a fermiophobic Higgs boson
   normalised to the fermiophobic cross section $\times$ branching ratio expectation as a function
   of $M_{H^{*}}$. The green and yellow bands indicate the corresponding $68$\% and $95$\% intervals.}
 \label{fig:fermiophobicHiggs}
\end{figure}

In the case of a fermiophobic Higgs $H^{*}$, couplings to fermions are forbidden, so that
production proceeds via VH and VBF only and the subsequent decays are also changed with respect
to the SM Higgs.
As a result, the fermiophobic Higgs decay $H^{*} \rightarrow \gamma\gamma$ branching ratio is
considerably enhanced with respect to the SM Higgs, as can be seen in
Figure~\ref{fig:fermiophobicHiggs}~(left).
A search by the ATLAS collaboration~\cite{atlasFermiophobicHiggs} observes a small excess in the
data of order $3.0\sigma$ in the same region as the SM Higgs search, as can be seen in
Figure~\ref{fig:fermiophobicHiggs}~(right).
With respect to the ATLAS SM Higgs search, the excluded region is extended down to Higgs
masses of $110$~GeV.
In a similar analysis by the CMS collaboration~\cite{cmsMoriondHiggs} an excess is also
observed at $126$~GeV, although when the $H^{*} \rightarrow WW$ and $H^{*} \rightarrow ZZ$
channels are also included, this is diluted to $1.0\sigma$ with masses $M_{H^{*}}< 190$~GeV
excluded at the $95$\% C.L.

%\clearpage
\section{Searches for physics beyond the Standard Model}
\label{sec:nonsusy}

A large variety of searches for BSM physics are performed in high energy physics, not
just at the LHC or the Tevatron, and not only within the Supersymmetry (SUSY) models
described in section~\ref{sec:susy}.
This includes searches for signatures such as: new $W'$ and $Z'$ bosons, heavy neutrinos
and $W_{R}$ production, large extra dimensions, narrow resonances in $2$ and $4$-jet spectra,
long lived particles and heavy stable charged particles, same sign leptons and black holes,
excited quarks and leptons, leptoquarks and contact interactions, searches for $t\bar{t}$ and
diboson resonances, heavy quarks and $4$th generation fermions.
A selection of these searches are described in the following.

%%%

\begin{figure}[t]
  \centering
  \includegraphics[width=0.57\textwidth]{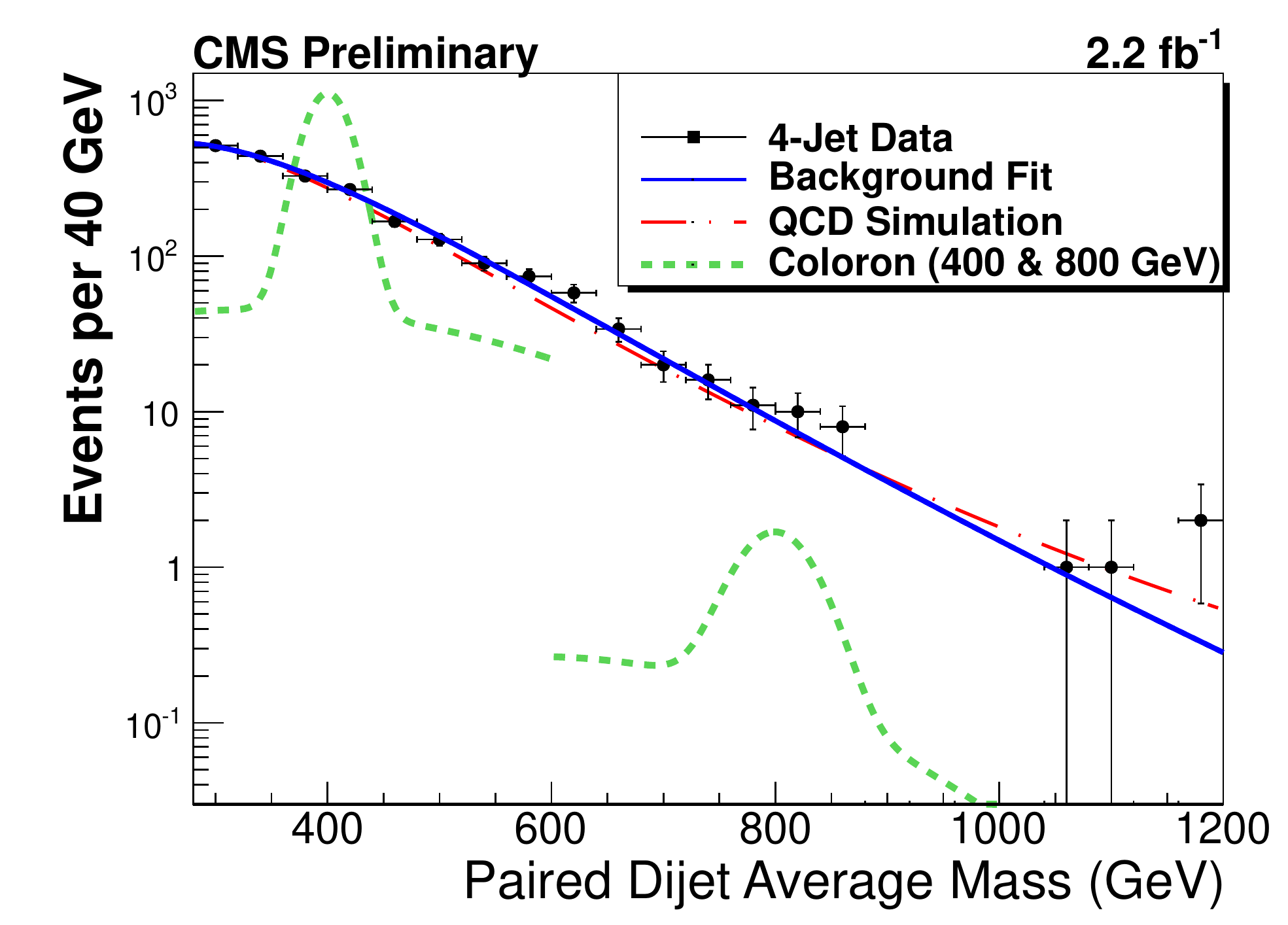}
  \hspace{0.1cm}
  \includegraphics[width=0.40\textwidth]{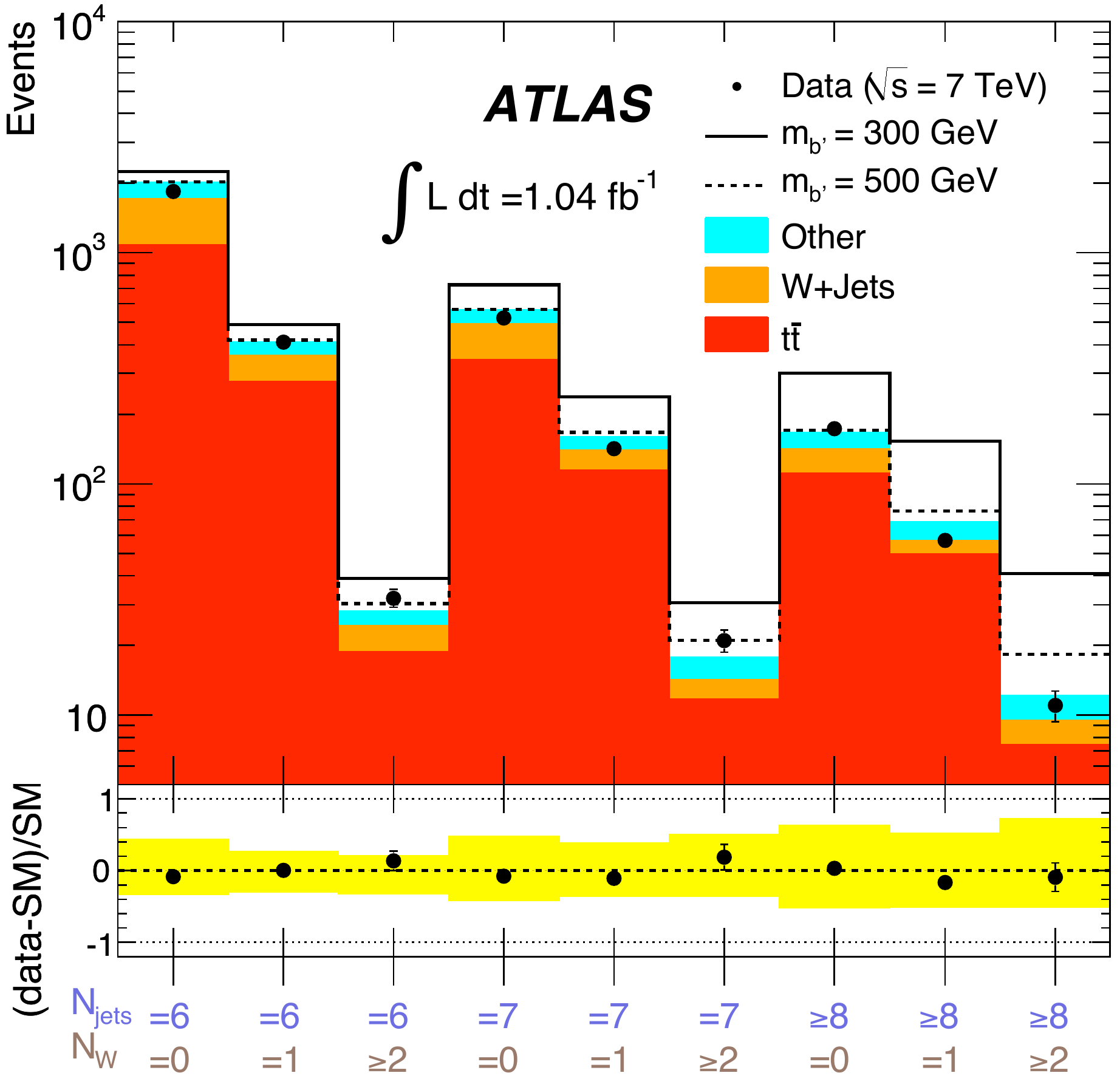}
  \caption{Left: The dijet mass distribution in the CMS data (black points) compared to a smooth
    background fit (solid curve) and a QCD MC based fit (dashed-dotted curve). Simulated Coloron
    resonances decaying to $q\bar{q}$ (green dashed curves) are also shown. Right:  A distribution of
    the number of events observed in the ATLAS data and that expected from SM processes for
    $N_{\rm jets} = 6, 7, >= 8$ and $N_{W} = 0, 1, >= 2$. The expected $b'$ signal for two masses is also
    shown, stacked on top of the SM background prediction.}
  \label{fig:nonsusy}
\end{figure}

Searches for new physics in events with leptons and/or jets are performed by the
ATLAS~\cite{policicchio} and CMS~\cite{weber} experiments. Among many limits, in an ATLAS
search for narrow resonances in dijet events, excited quarks are ruled out for masses
$M_{q^{*}} < 3.35$~TeV~\cite{atlasDijet}.
A similar CMS search also looks for new physics in $4$-jet final states, where the average dijet
mass spectrum is shown in Figure~\ref{fig:nonsusy} (left) and contains the expected signal
from the Coloron pair production model for two alternate mass scenarios~\cite{cmsDijetResonances}.
In the absence of a signal, limits are derived and Coloron masses in the range
$320 < M_{C} < 580$~GeV are ruled out at the $95$\% C.L.

%%%

Event topologies with many jets in addition to a lepton would result from $4$th generation
fermion pair production with multiple $W$ decays, and both CMS~\cite{chauhan} and ATLAS~\cite{zhong}
study such final states, requiring up to four $W$s with at least one decaying leptonically.
Figure~\ref{fig:nonsusy} (right) shows the number of events observed in the ATLAS analysis~\cite{ATLAS:2012aw},
where the data are divided into sub-samples depending on the number of jets and $W$s reconstructed
in the event.
A good agreement is observed between the data and the SM expectation, and a limit on
a $4$th generation quark mass $M_{b'} > 480$~GeV is set at the $95$\% C.L. 

%%%

Searches for new vector bosons are also performed at the LHC~\cite{policicchio,weber}, where a dilepton
search for a new $Z'$ boson by ATLAS (CMS) sets limits of $M_{Z'} > 2.21$ ($2.32$)~ TeV at the $95$\% C.L.
Searches for new $W'$ bosons now also explore a region in $M_{W'}^{T}$ beyond $2$~TeV.
A theoretical review was presented at the conference which included electroweak precision limits~\cite{salvioni}.

%%%

\begin{wrapfigure}{r}{0.54\textwidth}
 \centering
 \includegraphics[width=0.48\textwidth]{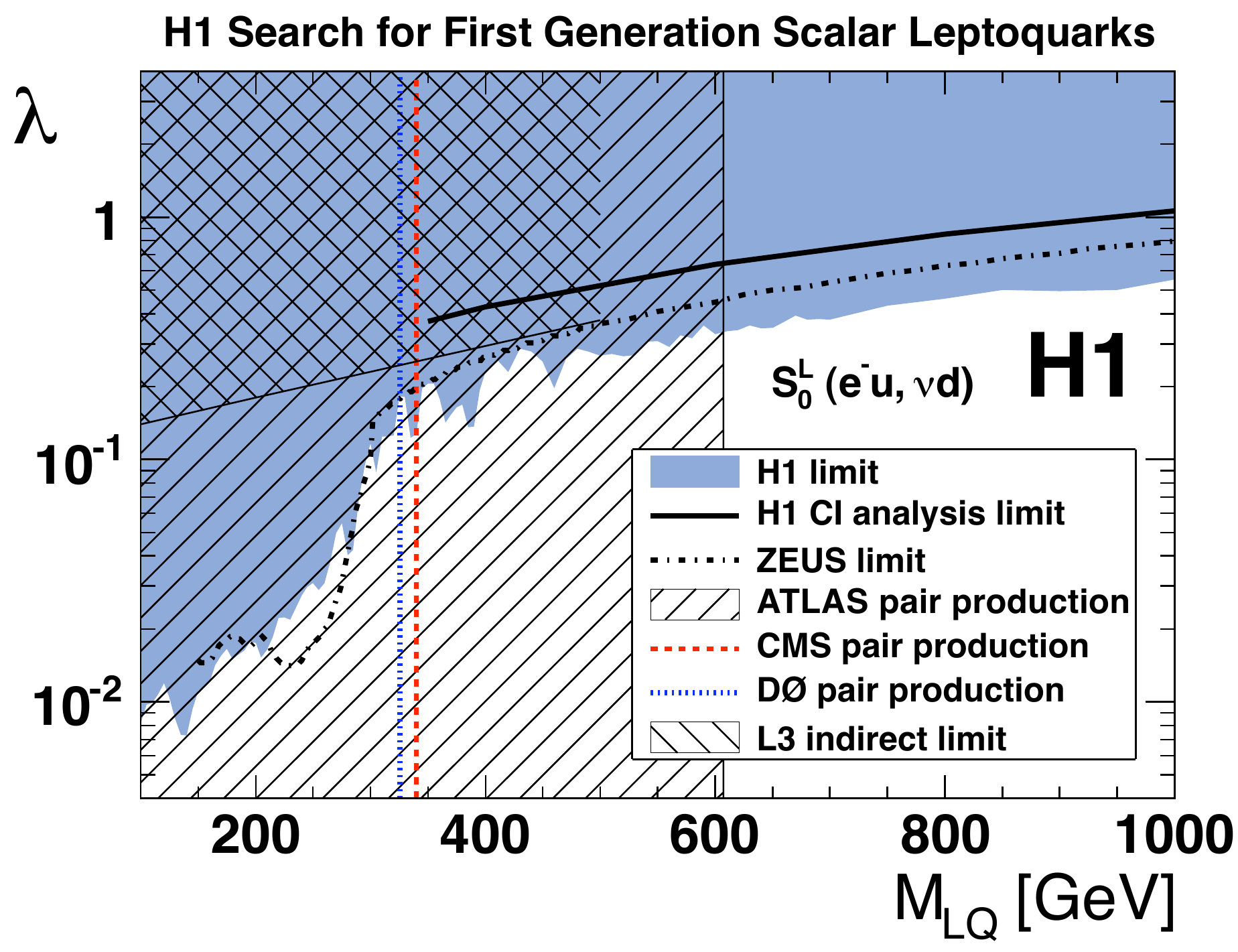}
 \caption{Exclusion limits on the coupling $\lambda$ as a function of leptoquark mass $M_{LQ}$
   for the $S^{L}_{1/2}$ ($\beta=0.5$) type leptoquark. The parentheses after the LQ name indicate the
   fermion pairs coupling to the LQ, where pairs involving anti-quarks are not shown. Domains above
   the curves and to the left of the horizontal lines are excluded at the $95$\% C.L.}
 \label{fig:lq}
\end{wrapfigure}
Complimentary searches for leptoquarks (LQs) are performed at hadron colliders, where LQs are produced
in pairs via the strong interaction and at HERA, where LQs are produced singly with a dependence on the
coupling $\lambda$.
Limits on the mass and coupling of a first generation LQ with an equal branching ratio to $eq$ and $\nu q$ ($\beta=0.5$)
are presented in Figure~\ref{fig:lq}.
The direct search limits from the ATLAS~\cite{policicchio}, CMS~\cite{chauhan} and D{\O}~\cite{d0lq} experiments are
compared to those from H1~\cite{pirumov} and ZEUS~\cite{antonelli}, as well as the indirect limits from the L3
experiment at LEP~\cite{l3lq}.
The most stringent limit from hadron colliders, $M_{LQ}>607$~GeV, is currently from ATLAS~\cite{atlaslq}, although for
large values of the coupling $\lambda$, the best limit is from H1~\cite{h1lq}.
The limit from an H1 contact interaction analysis~\cite{h1ci}, which is only sensitive to $LQ \rightarrow eq$ decays,
is also indicated in Figure~\ref{fig:lq}.

\clearpage
\section{Searches for Supersymmetry}
\label{sec:susy}

In this section, a review is presented of the results from ATLAS and CMS on recent SUSY searches with
many different final state channels, in addition to some theoretical points of view.
SUSY is motivated by possible solutions to the hierarchy problem, and may be within the reach of
the LHC experiments.
The largest SUSY production cross sections are expected to be from gluino and squark production,
and electroweak chargino and neutralino production may also be accessible.
At the LHC experiments, SUSY searches look for events with the following signatures: large MET,
hadronic activity, leptons (with different multiplicities from various channels), photons and heavy flavours.

%%%

$R-$parity distinguishes between SM particles ($R-$parity~$=+1$) and their SUSY partners
($R-$parity~$=-1$)~and if conserved protects the MSSM from rapid proton decay.
If $R-$parity is violated, then single SUSY particle production is possible and the lightest
supersymmetric particle (LSP) can decay further, meaning that MET searches (the mainstream
of SUSY searches) will not always be sensitive.
A search is performed by ATLAS within the $R-$parity violating mSUGRA model,
looking at $4$-lepton final states~\cite{meyer}, and excluding masses $M_{1/2} < 800$~GeV.
\begin{wrapfigure}{l}{0.60\textwidth}
  \centering
  \includegraphics[width=0.58\textwidth]{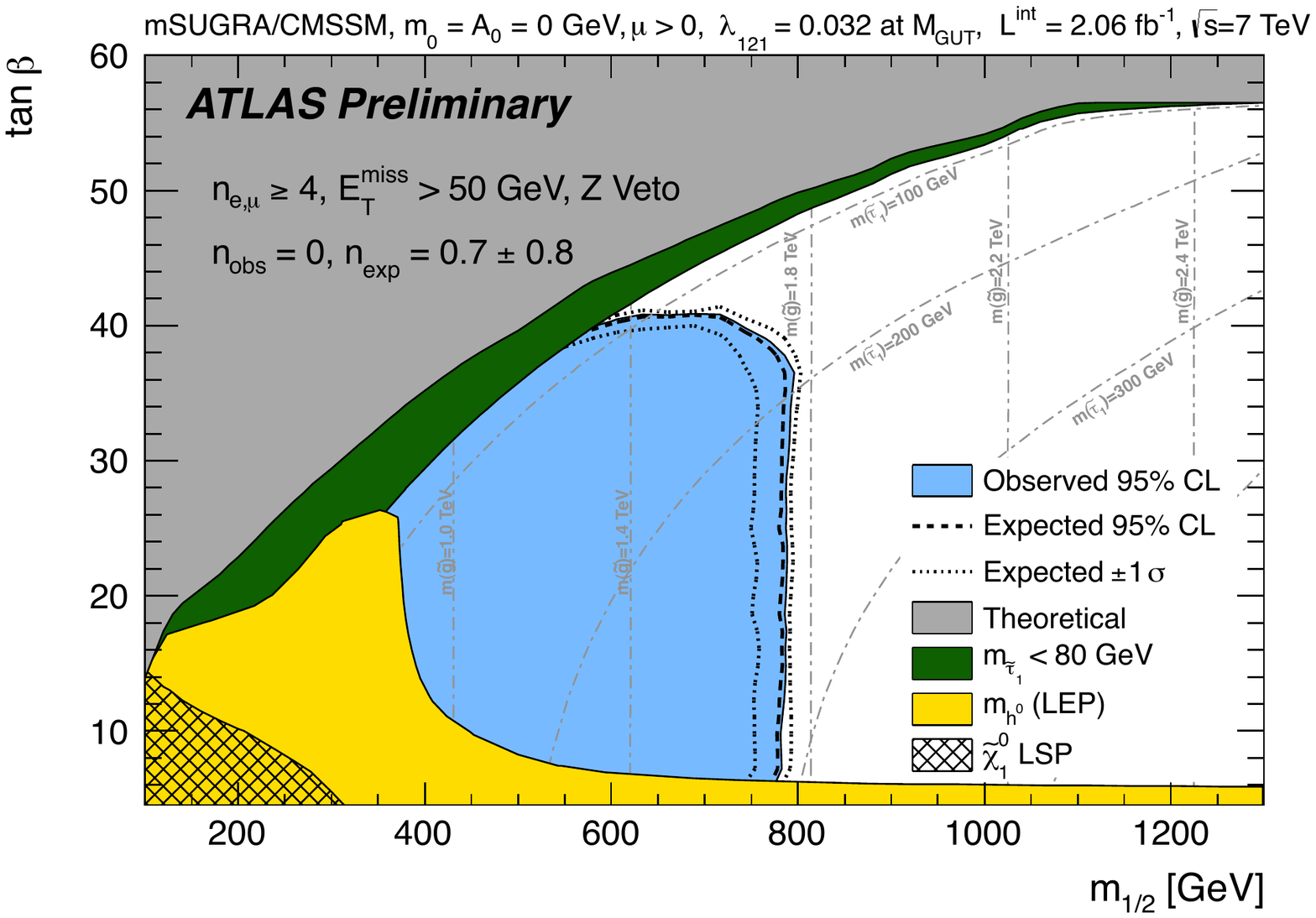}
  \caption{Limits at the $95$\% C.L from ATLAS in the mSUGRA/cMSSM m$_{1/2} - \beta$ plane from
    a search for Supersymmetry in final states with at least four isolated leptons and missing
    transverse momentum. The expected exclusion and its $\pm 1 \sigma$ variations are indicated by
    dashed lines. The other solid shaded areas are excluded by LEP results on the Higgs mass or
    because $m_{{\tilde{\tau}}_1} < 80$ GeV.}
  \label{fig:atlasRparityViolation}
\end{wrapfigure}
Figure~\ref{fig:atlasRparityViolation} shows the excluded region in the $M_{1/2} - \tan \beta$
plane at the $95$\% C.L. in a signal region defined as events with at least four leptons, MET $> 50$~GeV,
as well as a $Z$ boson veto $| M_{\ell} - M_{Z} | > 10$~GeV for all lepton pairs~\cite{atlasRParityViolation}. 
A search for strong $R-$parity conserving SUSY by ATLAS finds no evidence for a deviation from the SM
prediction and a limit for equal mass squarks and gluinos is set at around $1.4$~TeV~\cite{legger}.
A direct gaugino search is also performed by ATLAS, examining multi-lepton channels containing
two, three, or four or more leptons in combination with MET~\cite{heelan} and finds no statistically
significant evidence of SUSY.

%%%

\begin{figure}[htb]
  \centering
  \includegraphics[width=0.425\textwidth]{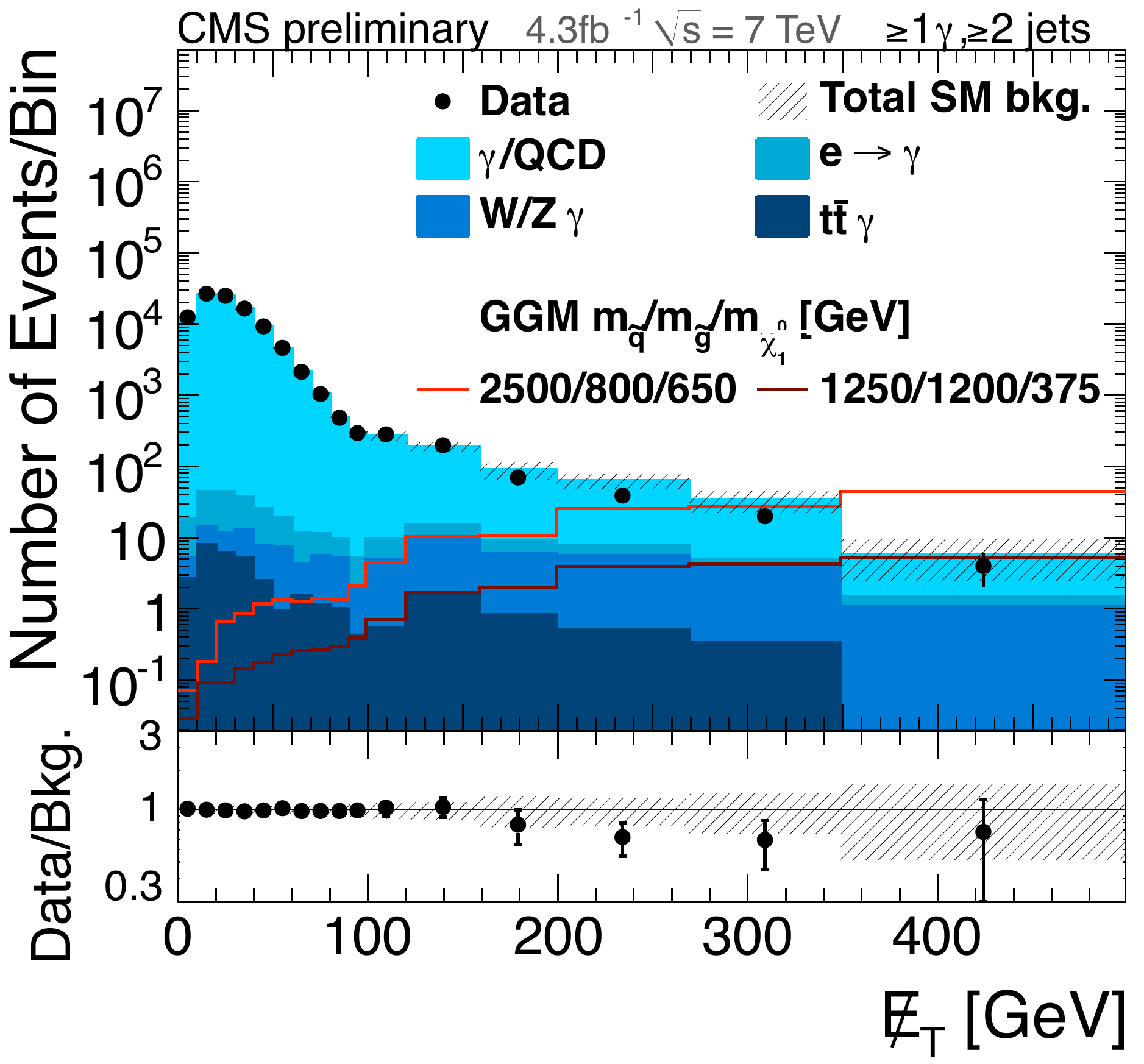}
  \includegraphics[width=0.565\textwidth]{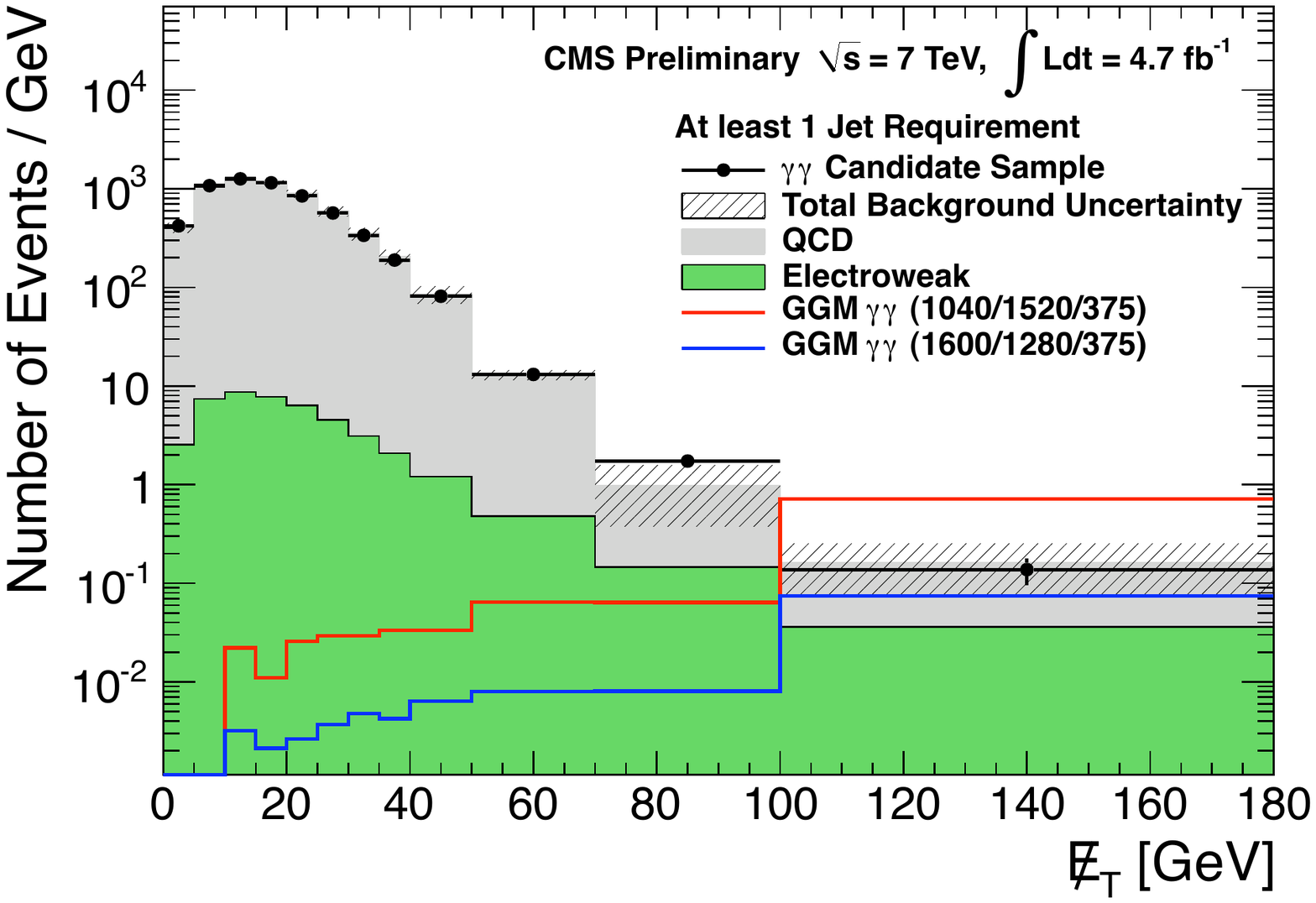}
  \caption{The missing transverse energy in the CMS single photon (left) and diphoton (right) SUSY
    analyses. The data events are compared to the SM background predictions, as well as
    the predictions from different general gauge mediation (GGM) signal benchmark points, with
    example squark/gluino/LSP masses as indicated in GeV.}
  \label{fig:cmsSUSYPhotons}
\end{figure}

CMS SUSY searches are categorised by different final states: jets$+$MET, jets$+$MET$+$leptons and 
photon decays from SUSY particles.
Jets$+$MET is a classical signature in SUSY searches and a large branching ratio is expected in this
channel.
Innovative new variables are employed to suppress the QCD background in this search: $M_{T2}$, which
is a transverse mass calculation in the case of two decay chains with missing particles and the Razor
variable $R=M_{R} / M^{R}_{T}$, which is an approximation of the scale and distribution
of the event~\cite{paktinat}.
No excess is seen in the data over the predicted SM backgrounds in the jets$+$MET search.
Jets$+$MET$+$lepton final states are also investigated by CMS, in which leptons arise primarily from neutralino
and chargino decays~\cite{niegel}.
Various analyses with single, dilepton, and multi-lepton final states are investigated and no excess above the SM
prediction is observed.
CMS also performs SUSY searches for final states with photons in addition to large MET (from gravitinos) and
multiple jets~\cite{jang}.
Both single photon and diphoton analyses observe no excess in the data compared to the SM prediction and
$95$\% C.L. upper limits on the cross section are set as $0.01$~pb for bino-like scenarios and $0.1$~pb 
for wino-like scenarios, where squark and gluino masses less than $1$~TeV are excluded.
Figure~\ref{fig:cmsSUSYPhotons} shows the MET distributions from the CMS single photon (left) and
diphoton (right) analyses~\cite{cmsSUSYPhotons}.

Third generation squarks can be lighter than the other generations.
This is particularly motivated by their role in solving the fine-tuning problem of the SM.
Stop and sbottom masses can still be only a few hundred GeV, providing the opportunity
for observing third generation squarks at the LHC.
Various searches are performed by the ATLAS and CMS experiments, looking for final states with
leptons, jets, MET, and $b$-jets, but so far no excess is observed over the SM
prediction~\cite{barajas,kalogeropoulos}. 

%%%

Three theory reviews were presented in the SUSY session at the conference.
The first two talks reviewed the status of the MSSM~\cite{bruemmer} and
beyond MSSM~\cite{lodone} scenarios after two years of LHC data taking.
Direct sparticle searches combined with a $125$~GeV Higgs give strong
constraints; the little hierarchy problem, constraints on neutralino dark matter
(DM), and heavy stops required by the $125$~GeV Higgs put some pressure on the
simplest models, although a natural SUSY spectrum can still be quite light.
The physics behind $Z'$ production in the MSSM was also presented~\cite{corcella}.
The $Z'$ may decay to SUSY particles such as sleptons, charginos, and neutralinos,
which then decay into SM particles.
In such a case, two or four leptons would appear in the final state, together with large
MET from the SUSY particles and the expected number of events with various scenarios
and branching ratios were discussed.

\section{Searches involving the top quark}
\label{sec:top}

The large mass of the top quark means that it has sizable coupling to the Higgs boson
and explains its special role in electroweak symmetry breaking.
Moreover, the experimental excess observed in the top pair ($t\bar{t}$) forward-backward asymmetry
$A_{FB}$ measurement from the Tevatron~\cite{tevatronAFB} may also hint at new physics.
A variety of searches involving the top quark are performed at the Tevatron and the LHC, which
is a top factory.
The top quark decays weakly followed by leptonic or hadronic $W$ decay, and therefore the signature
from top-pairs may be either dilepton, lepton$+$jets, or only jets in the final state.
The $t\bar{t}$ production cross section has been measured at the LHC using
$\sqrt{s}=7$~TeV data as $177~\pm~3~(\rm stat.)~^{+8}_{-7}~(\rm syst.)~\pm~7~(\rm lumi.)$~pb and
$165.8~\pm~2.2~(\rm stat.)~\pm~10.6~(\rm syst.)~\pm~7.8~(\rm lumi.)$~pb by the ATLAS ~\cite{atlasTopXs} and
CMS~\cite{cmsTopXs} experiments, respectively.

%%%

\begin{figure}[htb]
  \centering
  \includegraphics[width=0.48\textwidth]{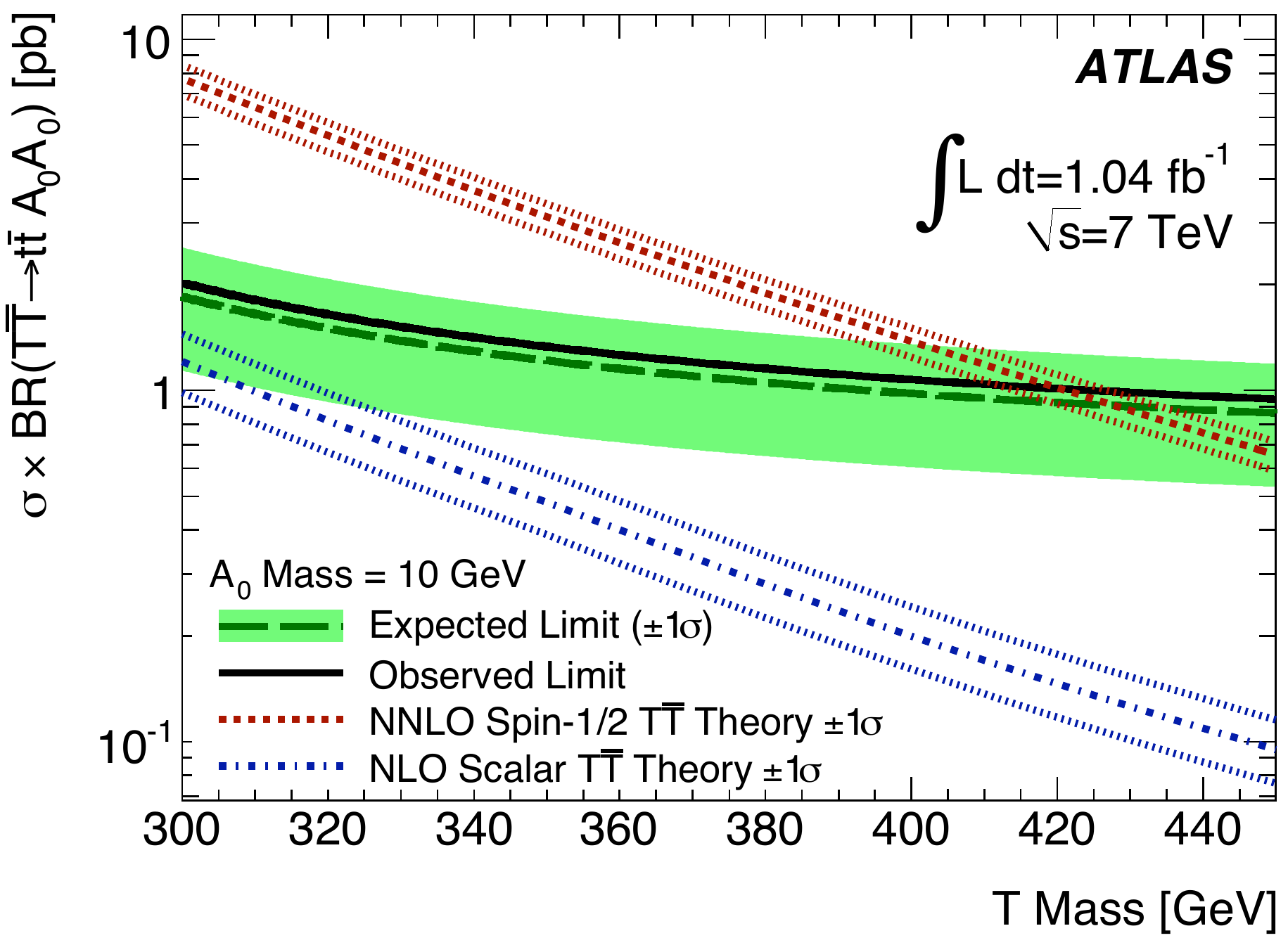}
  \includegraphics[width=0.48\textwidth]{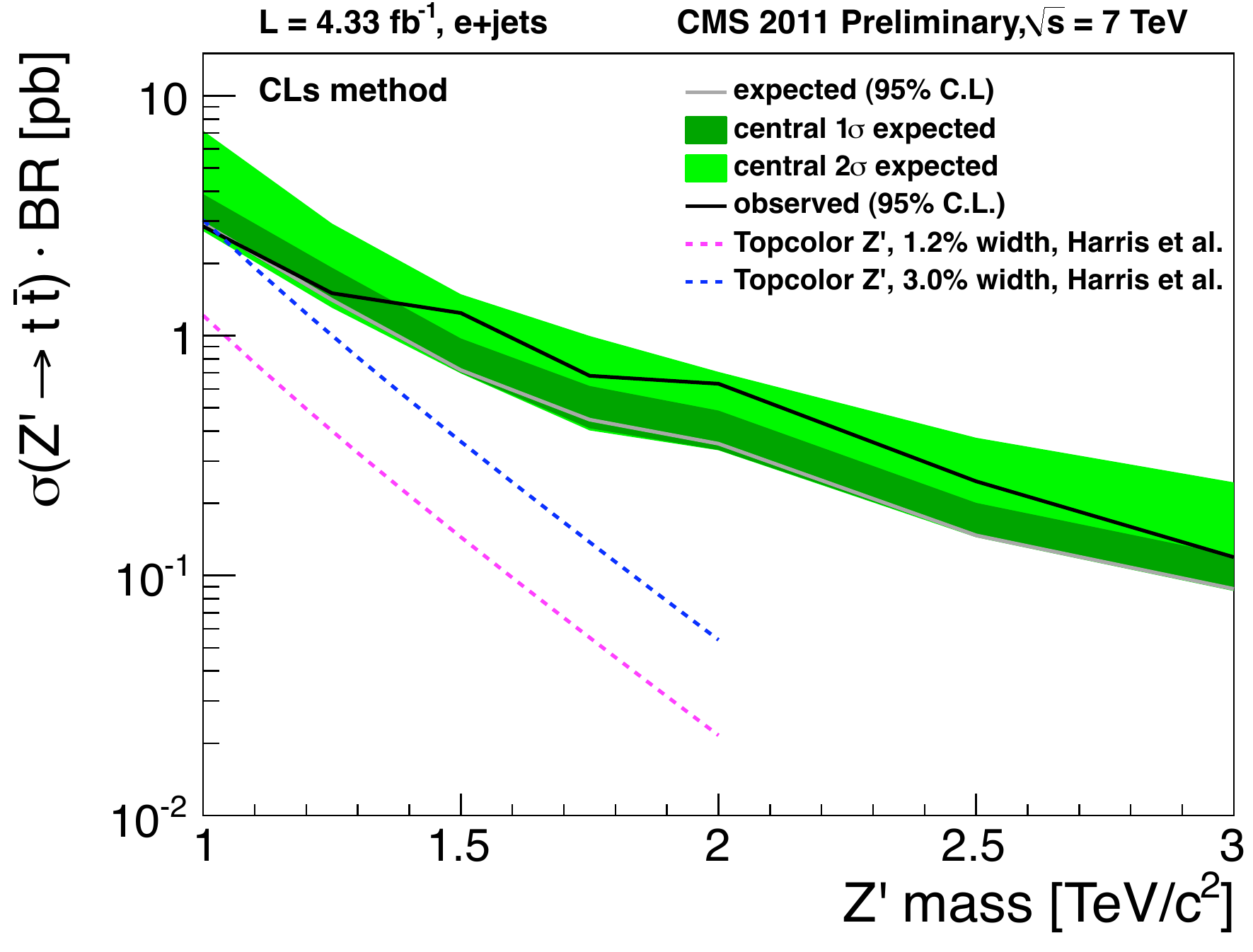}
  \caption{Left: Cross section $\times$ branching ratio excluded by ATLAS at the
    $95$\% C.L. as a function of the mass of a heavy quark like object $T$, for a scalar
    neutral mass $A_{0}=10$~GeV. Various theoretical predictions are also indicated.
    Right: Expected and observed upper limits at the $95$\% C.L. from CMS on the
    $\sigma(pp \rightarrow Z' \rightarrow t\bar{t})$ cross section $\times$
    branching ratio as a function of the $Z'$ mass. The expected signal from the
    Topcolor $Z'$ model is also shown.}
 \label{fig:top}
\end{figure}

A variety of top searches are performed by ATLAS looking for $t\bar{t}$ resonances, $t\bar{t}+$MET
final states and same sign top~\cite{calfayan}.
Whereas no evidence of new physics with top quarks is observed, a significant improvement of the
limits from the searches has been achieved.
CMS similarly performs searches for heavy resonances decaying to top pairs,
boosted tops, heavy bottom-like quarks and flavour changing neutral current (FCNC) in top
quark decays~\cite{bazterra}, where once again no evidence of new physics is observed.
Figure~\ref{fig:top} (left) shows the upper limits at the $95$\% C.L. from an ATLAS search for new
phenomena in $t\bar{t}+$MET final states~\cite{atlasTopSearches}.
Similar limits from a CMS search for high mass resonances decaying to $t\bar{t}$
in the electron$+$jets channel are shown in Figure~\ref{fig:top} (right).

Many top based searches for new physics are also performed by the CDF and D{\O}
experiments at the Tevatron, including among others: $t\bar{t}$ narrow resonances, top$+$jet
resonances, dark matter candidates associated with single top, anomalous couplings,
fourth generation quarks, boosted top quarks and Lorentz invariance violation.
No deviation from the SM prediction is reported in this comprehensive list of searches~\cite{peters}.

%%%

A search is performed by the ZEUS experiment for single top production in $ep$ collisions~\cite{antonelli}. 
Single top quark production at HERA has a cross section less than $1$~fb$^{-1}$, proceeding via the
charged current interaction.
However, single top may also be produced via the FCNC process, and a search is performed by ZEUS
based on the measurement of $W$ production, which examines the leptonic decay channels of the
$W$, with an additional top-like selection.
No excess above the SM prediction is observed in the data and constraints on the anomalous top
branching ratios $t \rightarrow u\gamma$ and $t \rightarrow uZ$ are established.

%%%

A theoretical review of the potential for new physics in the top sector was presented
at the conference, describing where significant hints of BSM physics can be found at
the LHC and the Tevatron~\cite{kamenik}.
For example, a large deviation from zero in the value of $A_{FB}$ could be due to
$s$-channel resonances at a mass scale of the order of a TeV.
At the LHC, this would manifest as an excess in the dijet and $t\bar{t}$ spectra, and such
analyses can provide constraints on BSM physics.
Further interesting possibilities of sub-TeV scale contributions in the $u$-channel or $t$-channel
are predicted at the LHC within the $t\bar{t}+$jets signature.

%%%

\section{$W$ and $Z$ physics}
\label{sec:wz}

\begin{figure}[t]
  \centering
  \includegraphics[width=0.50\textwidth]{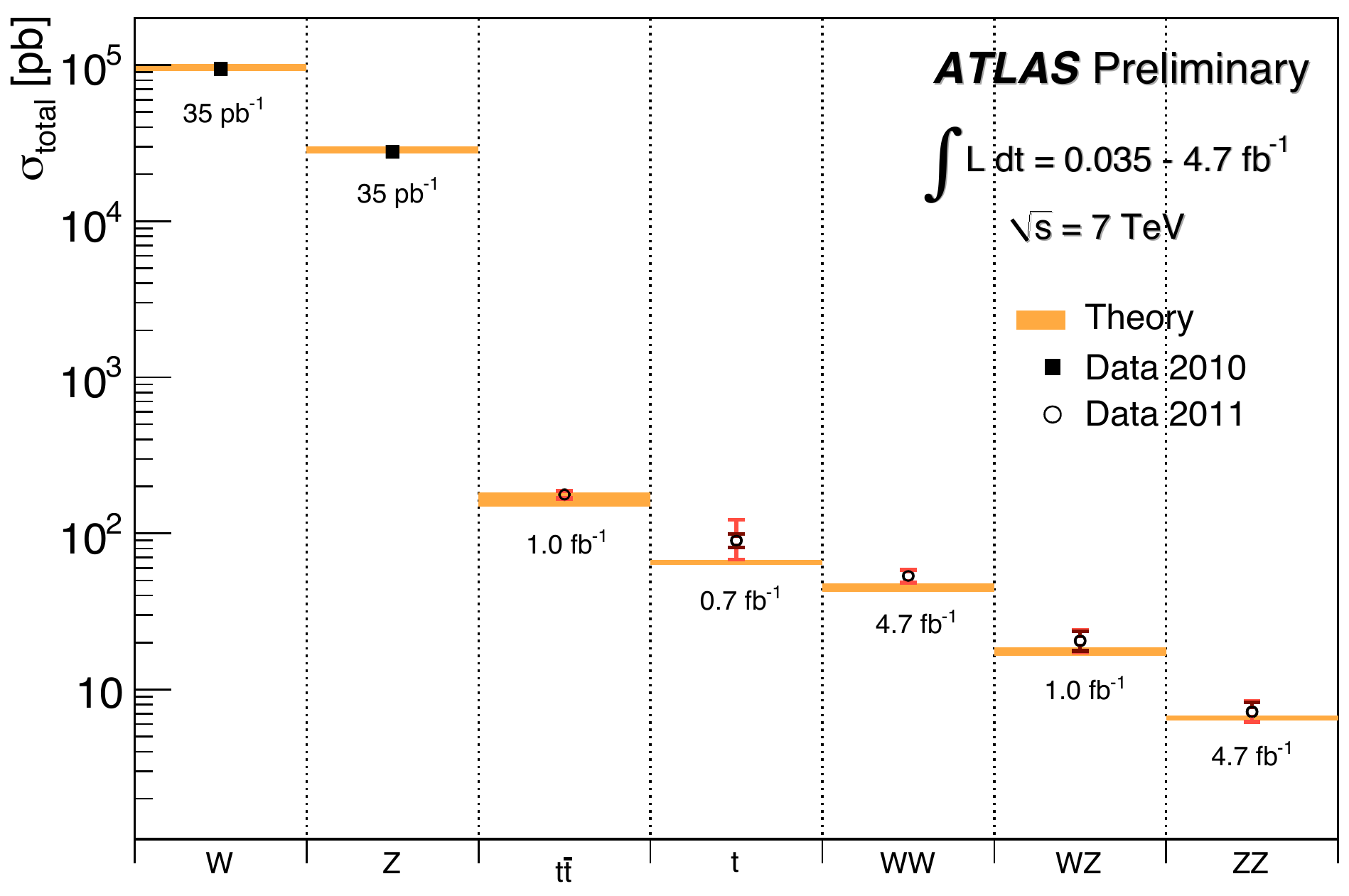}
  \hspace{0.7cm}
  \includegraphics[width=0.34\textwidth]{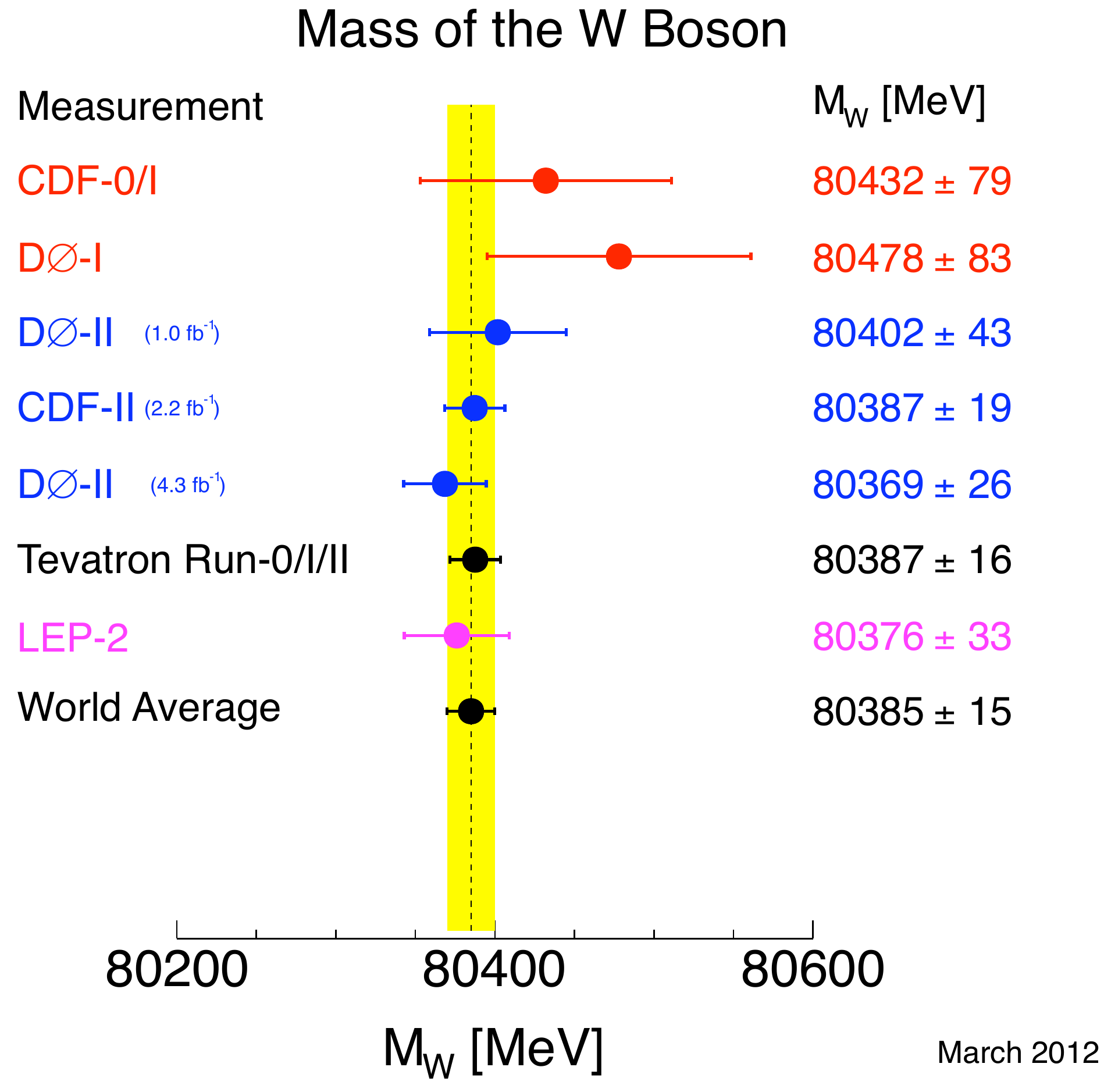}
  \caption{Left: A summary of the SM total production cross section measurements from the ATLAS
    experiment compared to the corresponding theoretical expectations. Right: A summary of
    measurements of the $W$ boson mass, including individual and combined Tevatron results, as well
    as the LEP measurement. An estimate of the world average, assuming no correlations between the
    Tevatron and LEP results, is also included in the figure.}
 \label{fig:w}
\end{figure}

Precision measurements with $W$ and $Z$ bosons not only provide sensitivity to new physics
but also serve as input to indirect searches for the Higgs.
Diboson cross section measurements are sensitive to triple gauge couplings (TGCs), allowing limits
on anomalous TGCs to be set.
In addition, the diboson process is an important background in Higgs searches, particularly in
the case of $WW$ and $ZZ$, and therefore the production cross section needs to be known precisely.
Such measurements are performed at the LHC by ATLAS~\cite{skottowe} and CMS~\cite{folgueras},
as well as at the Tevatron, where results from the CDF and D{\O} experiments are combined into one set of
measurements~\cite{vesterinen}.
SM cross section measurements from ATLAS are shown in Figure~\ref{fig:w} (left),
where a good agreement is observed with the SM predictions, which are calculated at NLO or higher.

%%%

A measurement of the $W$ mass gives indirect constraints on the Higgs mass and also provides 
a test of the SM.
If the LHC discovers the Higgs, one can compare indirect and direct mass measurements for
indications of BSM physics.
CDF and D{\O} recently updated their combined $W$ mass measurement using more data:
$2.2$~fb$^{-1}$ in the case of CDF and $4.3$~fb$^{-1}$ in the case of D{\O}~\cite{riddick}.
The measurement is performed using both electronic and muonic $W$ decays,
employing a template fitting method in the measurement.
The momentum scale calibration is an important part of the measurement in the muon channel
and is set by fits to $J/\psi, \Upsilon$ and $Z$ data.
In the electron channel, an energy scale calibration is applied: in the CDF measurement,
both $E/p$ calibrations from the momentum scale measurement and $M_{Z}$ measurement
are used to obtain a final energy scale, whereas in the D{\O} measurement, only $E/p$
calibration from the M$_{Z}$ measurement is used.
The CDF result is $M_{W} = 80387 \pm 19$~MeV and the D{\O} result $M_{W} = 80369 \pm 26$~MeV,
with a combined result of $M_{W} = 80387 \pm 16$~MeV, where the errors contain both
statistical and systematic uncertainties.
Figure~\ref{fig:w} (right) shows the summary of the measurements of the $W$ boson mass,
including the results from the Tevatron and their average~\cite{tevatronWMass},
as well as the measurement from LEP~\cite{lepWMass} and the resulting world average.

A measurement of elastic $Z$ production is performed by ZEUS using their complete $ep$ collision
data set~\cite{zeusElasticZ}. 
The SM cross section for $Z$ production in $ep$ collisions is expected to be much smaller than in $pp$,
due to the lack of $s$-channel Drell-Yan production.
The analysis examines the hadronic decay mode of the $Z$ in an elastic phase space, and a cross section of
$\sigma = 0.133^{+0.060}_{-0.057}~(\rm stat.)^{+0.049}_{-0.038}~(\rm syst.)$~pb is measured,
in good agreement with the SM prediction.

% References

{\raggedright
  \begin{footnotesize}
    
  \end{footnotesize}
}

\end{document}